\documentclass[12pt]{article}
\usepackage{amsmath,amssymb,graphicx,float}
\usepackage{braket}
\newcommand{\be}{\begin{equation}}
\newcommand{\bea}{\begin{eqnarray}}
\newcommand{\eea}{\end{eqnarray}}
\newcommand{\ba}{\begin{array}}
\newcommand{\ea}{\end{array}}
\newcommand{\ee}{\end{equation}}

\newcommand{\no}{\nonumber}

\newcommand{\dg}{\dagger}

\newcommand{\tr}{\mbox{Tr}}

\begin{document}

\begin{titlepage}

\vspace*{5mm}%

\title{\textbf {Two-Loop Integrability of ABJM Open Spin Chain from Giant Graviton}}
	\author{Nan Bai$^{a}$\footnote{bainan@mailbox.gxnu.edu.cn}~,Hui-Huang Chen$^{b}$\footnote{chenhh@jxnu.edu.cn}~,  Hao Ouyang$^{c}$\footnote{hao.ouyang@su.se}~,
		Jun-Bao Wu$^{d, e}$\footnote{junbao.wu@tju.edu.cn}}
	\date{}
\begin{flushright}\footnotesize
	
	\texttt{NORDITA-2019-001\\CJQS-2019-005} \\
	
\end{flushright}	
{\let\newpage\relax\maketitle}
	\maketitle
	\underline{}
	\vspace{-10mm}
	
	\begin{center}
		{\it
            $^{a}$ Department of Physics, Guangxi Normal University, \\Guilin 541004, China\\
			$^{b}$ College of Physics and Communication Electronics, Jiangxi Normal University, \\Nanchang 330022, China\\
			$^{c}$ Nordita, KTH Royal Institute of Technology and Stockholm University,
			Roslagstullsbacken 23, SE-106 91 Stockholm, Sweden\\
			$^{d}$ Center for Joint Quantum Studies and  Department of Physics, School of Science, Tianjin University, 135 Yaguan Road, Tianjin 300350, China\\
			$^{e}$ Center for High Energy Physics, Peking University, 5 Yiheyuan Road, Beijing 100871, China
		}
		\vspace{10mm}
	\end{center}

\begin{abstract}
  We prove the integrability of the two-loop open spin chain Hamiltonian from ABJM determinant like operators given in \cite{Chen:2018sbp}. By explicitly constructing R-matrices and K-matrices, we successfully obtain the two-loop Hamiltonian from the double row transfer matrices. This proves the integrability of our two-loop Hamiltonian. Based on the vacuum eigenvalues of the transfer matrices, we make a conjecture on the eigenvalues of the transfer matrices for general excited states. Bethe ansatz equations are simply obtained from the analytic conditions at the superficial poles of the eigenvalues.
\end{abstract}
\end{titlepage}
\section{Introduction}
\par Integrability found in $AdS_5/CFT_4$ correspondence enables us to study this conjecture in a non-trivially quantitative way. Tremendous progress has been achieved in solving planar $\mathcal{N}=4$ super Yang-Mills (SYM) using integrability techniques, see for example \cite{Beisert:2010jr, Basso:2015zoa}. ABJM theory is a three-dimensional $\mathcal{N}=6$  superconformal gauge theory dual to IIA string theory on $AdS_4\times \mathbb{CP}^3$ background \cite{Aharony:2008ug}.   Despite many results on integrable structure in this theory \cite{Klose:2010ki}, there are still some problems related to integrability in ABJM theory to be solved. For example, finding the full spectrum of open string attached on giant graviton  and determining the exact Bremsstrahlung functions using integrability in a way similar to  SYM case \cite{Bajnok:2012xc, Drukker:2012de, Correa:2012hh} remain important unsolved problems.\\
\par In \cite{Chen:2018sbp}, three of us obtained the two-loop open spin chain Hamiltonian from determinant like operators in ABJM theory. Our motivation for studying such kind of operators in ABJM theory came from the fact that these operators are dual to open strings ending on giant graviton in IIA string theory on $AdS_4\times \mathbb{CP}^3$ background \cite{Berenstein:2008dc,Giovannoni:2011pn, Lozano:2013ota, Cardona:2014ora} and a very similar setup in the $AdS_5/CFT_4$  context leads to an integrable open system \cite{Berenstein:2005vf, Hofman:2007xp}. Despite the strong evidence on integrability of this two-loop open spin chain provided in \cite{Chen:2018sbp} based on coordinate Bethe ansatz, we were lacking an explicit algebraic construction of Hamiltonian and  the corresponding Bethe ansatz equations at that time. \\
\par By algebraic construction, we mean that deriving the Hamiltonian from the commutating transfer matrices which serves as the generating function of an infinite number of conserved quantities. Our two-loop Hamiltonian is very special in the sense that the boundary degrees of freedom are not the same as the bulk ones, which makes it difficult to construct the double row transfer matrix. To encode this physical nature, we need the projected operator-valued $K$-matrices. By carefully choosing the c-number $K$-matrices, and using a similar method given in  \cite{Frahm:1998, Nepomechie:2011nz}, we successfully construct the projected operator-valued $K$-matrices which lead to the wanted boundary terms of the Hamiltonian. Our results establish a more solid ground for the two-loop integrability of the ABJM open spin chain.\\
\par The remaining part of this paper is organized as follows: In section~\ref{section2}, we review the two-loop Hamiltonian derived from the two-point functions of determinant like operators in ABJM theory and fix some useful notations. In section~\ref{section3}, we construct explicitly the projected operator-valued $K$-matrices and the double row transfer matrices, which generate our open spin chain Hamiltonian exactly. This establishes the integrable property of the Hamiltonian. In section~\ref{section4}, we derive Bethe ansatz equations from our ``guessed" eigenvalues of the double row transfer matrix. We also make some simple consistency checks for our proposal of the Bethe ansatz equations. Finally, we conclude in section~\ref{section5} and add three appendices to provide some computational details.
\section{Hamiltonian from determinant like operators in ABJM theory}\label{section2}
We consider the alternating spin chain model with open boundaries which originates from the anomalous dimension matrix of determinant like operators in ABJM theory. The state space is of the type
\bea\no
  \overset{\overset{\overset{{{A\!\!\!/}_1}}{\downarrow}}{1}}{\mathbb{C}}{}^{3}\otimes
  \overset{\overset{\overset{{\bar{\textbf{4}}}}{\downarrow}}{\bar{2}}}{\mathbb{C}}{}^{4}\otimes
  \overset{\overset{\overset{{{\textbf{4}}}}{\downarrow}}{3}}{\mathbb{C}}{}^{4}\otimes
  \cdots\otimes
  \overset{\overset{\overset{{\bar{\textbf{4}}}}{\downarrow}}{\overline{2L-2}}}{\mathbb{C}^4}\otimes
  \overset{\overset{\overset{{{\textbf{4}}}}{\downarrow}}{2L-1}}{\mathbb{C}^4}\otimes
  \overset{\overset{\overset{{{B\!\!\!/}_1}}{\downarrow}}{\overline{2L}}}{\mathbb{C}^3}.
  \eea
  The length of the spin chain is $2L$ with $2L-2$ 4-dimensional bulk spaces  plus two boundary spaces of 3 dimensions. The bulk consists of fundamental representation space of $SU(4)$ labeled as $\textbf{4}$ with the basis
  \bea
A_1=|1\rangle, A_2=|2\rangle,  B^{\dg}_1=|3\rangle, B^{\dg}_2=|4\rangle,
\eea
  and anti-fundamental representation space of $SU(4)$ labeled as $\bar{\textbf{4}}$ with the basis
  \bea
A^{\dg}_1=|1\rangle, A^{\dg}_2=|2\rangle,  B_1=|3\rangle, B_2=|4\rangle.
\eea
Note that, as explained in \cite{Chen:2018sbp}, the field $A_1$ does not exist at left boundary  and at right boundary $B_1$ is eliminated, so the dimensions at the boundaries are reduced to three.
\par The Hamiltonian of the spin chain is given by \cite{Chen:2018sbp}
\bea\label{Hamiltonian}
   H&=&\lambda^2 \sum_{l=1}^{L-2}\left(\mathbb{I}-\mathbb{P}_{2l+1, 2l+3}+\frac12\mathbb{P}_{2l+1, 2l+3}\mathbb{K}_{2l+1, \overline{2l+2}}+\frac12\mathbb{P}_{2l+1, 2l+3}\mathbb{K}_{\overline{2l+2}, 2l+3}\right)
   Q^{A_1}_1Q^{B_1}_{\overline{2L}} \\ \no
   &+&\lambda^2\sum_{l=1}^{L-2}\left(\mathbb{I}-\mathbb{P}_{\overline{2l},\overline{2l+2}}+\frac{1}{2}\mathbb{P}_{\overline{2l},\overline{2l+2}}\mathbb{K}_{\overline{2l},2l+1}
   +\frac{1}{2}\mathbb{P}_{\overline{2l},\overline{2l+2}}\mathbb{K}_{2l+1,\overline{2l+2}}\right)Q^{A_1}_1Q^{B_1}_{\overline{2L}}\\\no
  &+&\lambda^2Q_1^{A_1}\left(\mathbb{I}+\frac{1}{2}\mathbb{K}_{1,\bar{2}}-\mathbb{P}_{1,3}+\frac{1}{{2}}\mathbb{P}_{1,3}\mathbb{K}_{1,\bar{2}}+\frac{1}{2}\mathbb{P}_{1,3}\mathbb{K}_{\bar{2},3}\right)Q_1^{A_1}Q^{B_1}_{\overline{2L}}\\\no
  &+&\lambda^2Q^{A_1}_1Q^{B_1}_{\overline{2L}}\left(\mathbb{I}+\frac{1}{2}\mathbb{K}_{2L-1,\overline{2L}}-\mathbb{P}_{\overline{2L-2},\overline{2L}}+\frac{1}{2}\mathbb{P}_{\overline{2L-2},\overline{2L}}\mathbb{K}_{\overline{2L-2},2L-1}
  +\frac{1}{2}\mathbb{P}_{\overline{2L-2},\overline{2L}}\mathbb{K}_{2L-1,\overline{2L}}\right)Q^{B_1}_{\overline{2L}}\\\no
  &+&\lambda^2 Q^{A_1}_1\left(\mathbb{I}-Q_{\bar{2}}^{A^{\dg}_1}\right)Q^{B_1}_{\overline{2L}}+\lambda^2 Q^{A_1}_1\left(\mathbb{I}-Q_{2L-1}^{B^{\dg}_1}\right)Q^{B_1}_{\overline{2L}}
  \eea
  where ``~i~" and ``~$\bar{i}~$" denote the fundamental and anti-fundamental representation spaces of $SU(4)$ respectively and $\lambda\equiv{N}/{k}$ is the 't~Hooft coupling constant of ABJM theory. The projector $Q$ is defined as
  \bea
  Q^{X}|X\rangle=0,\qquad Q^{X}|Y\rangle=|Y\rangle \quad \mbox{for} \quad X\neq Y.
  \eea
  In the standard basis, the two projectors have the matrix form
  \bea
  Q^{A_1}=
  \begin{pmatrix}
  0&&&\\
  &1&&\\
  &&1&\\
  &&&1
  \end{pmatrix},\quad
  Q^{B_1}=
  \begin{pmatrix}
  1&&&\\
  &1&&\\
  &&0&\\
  &&&1
  \end{pmatrix}.
  \eea
  The operators $\mathbb{P}$ and $\mathbb{K}$ are defined by the standard elementary matrices $e_{ab}$ (with components $[e_{ab}]_{ij}=\delta_{ai}\delta_{bj}$) as
\bea
\mathbb{P}=\sum_{a,b=1}^4 e_{ab}\otimes e_{ba},\quad \mathbb{K}=\sum_{a,b=1}^4 e_{ab}\otimes e_{ab}.
\eea
\section{Integrable constructions of the Hamiltonian}\label{section3}
In this section, we will demonstrate the integrability of the Hamiltonian (\ref{Hamiltonian}) by the algebraic approach based on the quantum inverse scattering method for open boundary \cite{Sklyanin:1988yz} and the projecting method \cite{Frahm:1998}.
\subsection{General procedure}
For the alternating spin chain model with the origin of ABJM theory, we adopt the following four R-matrices \cite{Minahan:2008hf, Bak:2008cp},
\bea
&&R_{12}(u)=u+\mathbb{P}_{12},\\
&&R_{\bar1\bar2}(u)=u+\mathbb{P}_{\bar1\bar2},\\
&&R_{1\bar2}(u)=-(u+2)+\mathbb{K}_{1\bar2},\\
&&R_{\bar12}(u)=-(u+2)+\mathbb{K}_{\bar12}.
\eea
These R matrices satisfy the following Yang-Baxter equations,
\bea
R_{12}(u-v)R_{13}(u)R_{23}(v)&=&R_{23}(v)R_{13}(u)R_{12}(u-v),\\
R_{12}(u-v)R_{1\bar{3}}(u)R_{2\bar{3}}(v)&=&R_{2\bar{3}}(v)R_{1\bar{3}}(u)R_{12}(u-v),\\
R_{1\bar{2}}(u-v)R_{13}(u)R_{\bar{2}3}(v)&=&R_{\bar{2}3}(v)R_{13}(u)R_{1\bar{2}}(u-v),\\
R_{\bar{1}2}(u-v)R_{\bar{1}3}(u)R_{23}(v)&=&R_{23}(v)R_{\bar{1}3}(u)R_{\bar{1}2}(u-v),
\eea
and are symmetric
\bea
\mathbb{P}_{12}R_{12}(u)\mathbb{P}_{12}&=&R_{12}(u),\\
\mathbb{P}_{\bar{1}2}R_{\bar{1}2}(u)\mathbb{P}_{\bar{1}2}&=&R_{\bar{1}2}(u),
\eea
\bea
R_{12}(u)^{t_1}&=&R_{12}(u)^{t_2},\\
R_{\bar{1}2}(u)^{t_{\bar{1}}}&=&R_{\bar{1}2}(u)^{t_2},
\eea
where the partial transpose operator $t_i$ acts on the space $V_i$.
Unitarity and crossing unitarity are also satisfied,
\bea
R_{12}(u)R_{12}(-u)&=&1-u^2,\\
R_{\bar{1}2}(u)R_{\bar{1}2}(-u)&=&4-u^2,
\eea
\bea
R_{12}(u)^{t_1}R_{12}(-u-4)^{t_1}&=&u(-u-4),\\
R_{\bar{1}2}(u)^{t_{\bar{1}}}R_{\bar{1}2}(-u-4)^{t_{\bar{1}}}&=&-u^2-4u-3.
\eea

In order to deal with the open boundary systems, the reflection K-matrices are needed. For the trivial boundaries without dynamics, these K-matrices are the c-number solutions of the reflection equations (REs) shown below\footnote{Precisely speaking, these two reflection matrices $K_1(u)$ and $K_{\bar{1}}(u)$ will generate the right boundary terms, while for the left boundary, we need another set of two reflection matrices satisfying the so-called dual reflection equations\cite{Sklyanin:1988yz}. The discussions on the left boundary are postponed to subsection~\ref{subsection33}. Also notice that though (\ref{RE2}) and (\ref{RE4}) can  be obtained from (\ref{RE1}) and (\ref{RE3}) using conjugation symmetry, the solutions we need  do not respect this symmetry, see (\ref{cs1}, \ref{cs2}). },
\bea
&&R_{12}(u-v)K_1(u)R_{21}(u+v)K_{2}(v)=K_{2}(v)R_{12}(u+v)K_1(u)R_{21}(u-v), \label{RE1}\\
&&R_{\bar1\bar2}(u-v)K_{\bar1}(u)R_{\bar{2}\bar{1}}(u+v)K_{\bar{2}}(v)=K_{\bar{2}}(v)R_{\bar{1}\bar{2}}(u+v)K_{\bar{1}}(u)R_{\bar{2}\bar{1}}(u-v),\label{RE2}\\
&&R_{1\bar{2}}(u-v)K_{1}(u)R_{\bar{2}1}(u+v)K_{\bar{2}}(v)=K_{\bar{2}}(v)R_{1\bar2}(u+v)K_{1}(u)R_{\bar{2}1}(u-v),\label{RE3}\\
&&R_{\bar{1}2}(u-v)K_{\bar{1}}(u)R_{2\bar{1}}(u+v)K_2(v)=K_2(v)R_{\bar{1}2}(u+v)K_{\bar{1}}(u)R_{2\bar{1}}(u-v).\label{RE4}
\eea
If the open system contains dynamic particles or impurities at the boundary, it is worth trying a kind of operator-valued K-matrices called projected K-matrices \cite{Frahm:1998}. Here we will give a brief review of the prescriptions of the projecting method:
\par 1. First we start from a carefully chosen c-number solution $K(u)$ of the RE.
\par 2. Then we define an operator-valued matrix $\mathcal{K}^-(u)$ as
\bea
\mathcal{K}^-_{0i}(u)=R_{0i}(u)K_0(u)R^{-1}_{0i}(-u)
\eea
where we have explicitly labeled the auxiliary space $V_0$ and the internal space $\mathcal{H}\equiv V_i$ as ``0" and ``i" respectively in the subscript of $\mathcal{K}^-(u)$. Note that the internal space $\mathcal{H}$ introduced by the R-matrix is the place where the boundary degrees of freedom reside.  This $\mathcal{K}^-(u)$ will satisfy the RE by construction and usually be called the regular solution.
\par 3. If the boundary internal space  is  smaller than the bulk state space, i.e. dim $\mathcal{H} < $ dim $V_i$, then we can decompose the total space $V_i$ into $\mathcal{H}$ and its orthogonal complement $\mathcal{H}^{\perp}$ by two projectors $Q$ and $Q^\perp$ respectively, such that $V_i=\mathcal{H}\oplus \mathcal{H}^{\perp}$. An important observation here is that if either of the following two projections vanish,
\bea
Q_i\mathcal{K}_{0i}^-(u)Q_i^\perp=0,\quad \mbox{or}\quad Q_i^\perp\mathcal{K}_{0i}^-(u)Q_i=0,\label{projection}
\eea
then the projections $Q_i\mathcal{K}_{0i}^-(u)Q_i$ and $Q_i^\perp\mathcal{K}_{0i}^-(u)Q_i^\perp$ will solve the RE. The new solutions of RE obtained in this way are called the projected K-matrices \cite{Frahm:1998}.\footnote{The projected K-matrices give the so-called 'singular' boundary matrices in \cite{Zhou1, Zhou2} which were introduced to  treat the situations with smaller boundary internal space.} We need both $\mathcal{K}^-_{0i}(u)$ and $\mathcal{K}^-_{\bar{0}i}(u)$ with auxiliary space $V_0$ ($V_{\bar{0}}$) being the $\textbf{4}$ ($\bar{\textbf{4}}$) representation.
\par As for the projected $K^+(u)$ matrices corresponding to the left boundary terms, we can find them by the isomorphic map from projected $K^-(u)$ matrices \cite{Sklyanin:1988yz}, the details will be provided later. With the above R-matrices and projected K-matrices, we can construct the following two double row transfer matrices for the alternating spin chain with dynamic boundaries,
\bea
\tau(u)=\tr_0K^{+}_{01}(u)R_{0\bar2}(u)\cdots R_{0,2L-1}(u)K^{-}_{0,\overline{2L}}(u)R_{0,2L-1}(u)\cdots R_{0\bar2}(u),\\
\bar{\tau}(u)=\tr_{\bar{0}}K^{+}_{\bar{0}1}(u)R_{\bar{0}\bar2}(u)\cdots R_{\bar{0},2L-1}(u)K^-_{\bar{0},\overline{2L}}(u)R_{\bar{0},2L-1}(u).\cdots R_{\bar{0}\bar2}(u).
\eea
Then it can be shown  that the transfer matrices obey the commutativity property
\begin{equation}
  [\bar\tau(u),\bar\tau(v)]=[\tau(u),\bar\tau(v)]=[\tau(u),\tau(v)]=0.
\end{equation}
The Hamiltonian can be obtained from the transfer matrices by
\bea
\mathcal{H}=\frac{d}{du}\log\tau(u)\bigg|_{u=0}+\frac{d}{du}\log{\bar{\tau}(u)}\bigg|_{u=0}.
\eea
\subsection{Projected $K^-(u)$ matrices and the right boundary terms}
Given any projected $K^-$-matrix, we will have the related right boundary term. For our alternating spin chain, the following two projected $K^-(u)$ matrices are needed,
\bea
  &&K^-_{1\bar{2}}(u)=Q^{B_1}_{\bar{2}}R_{1\bar{2}}(u)K_{1}(u)R^{-1}_{1\bar{2}}(-u)Q^{B_1}_{\bar{2}},\label{singular}\\\no
  &&K^-_{\bar{1}\bar{2}}(u)=Q^{B_1}_{\bar{2}}R_{\bar{1}\bar{2}}(u)K_{\bar{1}}(u)R^{-1}_{\bar{1}\bar{2}}(-u)Q^{B_1}_{\bar{2}},
  \eea
  where $K_{1}(u)$ and $K_{\bar{1}}(u)$ are two c-number matrices to be determined later. By construction, the projected $K^-$-matrix satisfy the ``operator" reflection equations,
  \bea
&&R_{12}(u-v)K^-_{1\bar3}(u)R_{21}(u+v)K^-_{2\bar3}(v)=K^-_{2\bar3}(v)R_{12}(u+v)K^-_{1\bar3}(u)R_{21}(u-v), \label{ORE1}\\
&&R_{\bar1\bar2}(u-v)K^-_{\bar1\bar3}(u)R_{\bar{2}\bar{1}}(u+v)K^-_{\bar{2}\bar3}(v)=K^-_{\bar{2}\bar3}(v)R_{\bar{1}\bar{2}}(u+v)K^-_{\bar{1}\bar3}(u)R_{\bar{2}\bar{1}}(u-v),\\
&&R_{1\bar{2}}(u-v)K^-_{1\bar3}(u)R_{\bar{2}1}(u+v)K^-_{\bar{2}\bar3}(v)=K^-_{\bar{2}\bar3}(v)R_{1\bar2}(u+v)K^-_{1\bar3}(u)R_{\bar{2}1}(u-v),\\
&&R_{\bar{1}2}(u-v)K^-_{\bar{1}\bar3}(u)R_{2\bar{1}}(u+v)K^-_{2\bar3}(v)=K^-_{2\bar3}(v)R_{\bar{1}2}(u+v)K^-_{\bar{1}\bar3}(u)R_{2\bar{1}}(u-v).\label{ORE4}
\eea
  Formally, the projected $K^-$-matrices will produce the right boundary terms as follows,
  \bea\label{rbt}
  \mathcal{H}^{total}_{r}&=&\left[K^{-}_{2L-1,\overline{2L}}(0)\right]^{-1}\left[\frac{d}{du}K^{-}_{2L-1,\overline{2L}}(u)\bigg|_{u=0}\right]\\\no
  &+&[-2+\mathbb{K}_{\overline{2L-2},2L-1}]^{-1}\left[K^{-}_{\overline{2L-2},\overline{2L}}(0)\right]^{-1}\left[\frac{d}{du}K^{-}_{\overline{2L-2},\overline{2L}}(u)\bigg|_{u=0}\right][-2+\mathbb{K}_{\overline{2L-2},2L-1}].
  \eea
  We provide the details of the calculations in the appendix~A.
\subsection{Projected $K^+(u)$ matrices and the left boundary terms}\label{subsection33}
The projected $K^+$-matrices account for the left boundary terms of the Hamiltonian and could be obtained by the isomorphism from $K^-$. In particular, for the symmetric boundary terms, i.e. the ones having the same expressions but acting on different state spaces, the Sklyanin's ``less obvious'' Z-isomorphism \cite{Sklyanin:1988yz} is the most suitable choice. In our case, we use the following two projected $K^+$-matrices,
\bea
&&K^+_{13}(u)=k(u)\tr_2\mathbb{P}_{12}R_{12}(-2u-4)K^-_{23}(u),\label{kk1}\\
&&K^+_{\bar{1}3}(u)=\bar{k}(u)\tr_{\bar{2}}\mathbb{P}_{\bar{1}\bar{2}}R_{\bar{1}\bar{2}}(-2u-4)K^-_{\bar{2}3}(u).\label{kk2}
\eea
where $k(u)$ and $\bar{k}(u)$ are two arbitrary scalar functions and the projected $K^-$-matrices are given by
\bea
&&K^-_{12}(u)=Q^{A_1}_2R_{12}(u)\tilde{K}_1(u)R^{-1}_{12}(-u)Q^{A_1}_2,\\
&&K^-_{\bar{1}2}(u)=Q^{A_1}_2R_{\bar{1}2}(u)\tilde{K}_{\bar{1}}(u)R^{-1}_{\bar{1}2}(-u)Q^{A_1}_2.
\eea
Note that we have used two new c-number solutions $\tilde{K}_{1}(u)$ and $\tilde{K}_{\bar{1}}(u)$ to formulate the projected $K^+$-matrices. By construction, the projected $K^+$-matrix satisfy the ``operator"  dual reflection equations,
\bea
R_{12}(-u+v)K^+_{13}(u)^{t_1}R_{21}(-u-v-4)K^+_{23}(v)^{t_2}\\\no
=K^+_{23}(v)^{t_2}R_{12}(-u-v-4)K^+_{13}(u)^{t_1}R_{21}(-u+v),
\eea
\bea
R_{\bar1\bar2}(-u+v)K^+_{\bar13}(u)^{t_{\bar1}}R_{\bar2\bar1}(-u-v-4)K^+_{\bar23}(v)^{t_{\bar2}}\\\no
=K^+_{\bar23}(v)^{t_{\bar2}}R_{\bar1\bar2}(-u-v-4)K^+_{\bar13}(u)^{t_{\bar1}}R_{\bar2\bar1}(-u+v),
\eea
\bea
R_{1\bar2}(-u+v)K^+_{13}(u)^{t_1}R_{\bar21}(-u-v-4)K^+_{\bar23}(v)^{t_{\bar2}}\\\no
=K^+_{\bar23}(v)^{t_{\bar2}}R_{1\bar2}(-u-v-4)K^+_{13}(u)^{t_1}R_{\bar21}(-u+v),
\eea
\bea
R_{\bar12}(-u+v)K^+_{\bar13}(u)^{t_{\bar1}}R_{2\bar1}(-u-v-4)K^+_{23}(v)^{t_2}\\\no
=K^+_{23}(v)^{t_2}R_{\bar12}(-u-v-4)K^+_{\bar13}(u)^{t_{\bar1}}R_{2\bar1}(-u+v).
\eea
\par The projected $K^+$-matrices will produce the following kind of left boundary terms
\bea\label{lbt}
\mathcal{H}^{total}_l&=&\left[\tr_0K^+_{01}(0)\right]^{-1}\left\{\tr_0\frac{dK^+_{01}(u)}{du}\bigg|_{u=0}+
\tr_0\left(K^+_{01}(0)\left(\mathbb{I}+2\mathbb{P}_{03}-\mathbb{P}_{03}\mathbb{K}_{0\bar2}-\mathbb{K}_{0\bar2}\mathbb{P}_{03}\right)\right)\right\}\\\no
&+&\left[\tr_{\bar{0}}K^+_{{\bar{0}}1}(0)\right]^{-1}\left(\tr_{\bar{0}}\frac{dK^+_{{\bar{0}}1}(u)}{du}\bigg|_{u=0}+
2\tr_{\bar{0}}\left(K^{+}_{\bar{0}1}(0)\mathbb{P}_{\bar{0}\bar2}\right)\right).
\eea
More details about the calculations can be found in the Appendix~A.
\subsection{Suitable c-number solutions and the integrability of the model}
We claim that the needed c-number solutions of the reflection equations for projected $K^-$-matrices are
\bea
K_0(u)=
\begin{pmatrix}
1+u&&&\\
&1+u&&\\
&&1-u&\\
&&&1+u
\end{pmatrix}=\left(2Q_0^{B_1^\dag}-1\right)u+1 \label{cs1}
\eea
and
\bea
K_{\bar{0}}(u)=
\begin{pmatrix}
1&&&\\
&1&&\\
&&-1&\\
&&&1
\end{pmatrix}=2Q_{\bar{0}}^{B_1}-1.\label{cs2}
\eea
 For projected $K^+$-matrices, the c-number solutions are given by
\bea
\tilde{K}_{0}(u)=
\begin{pmatrix}
-1&&&\\
&1&&\\
&&1&\\
&&&1
\end{pmatrix}=2Q_0^{A_1}-1 \label{cs3}
\eea
and
\bea
\tilde{K}_{\bar{0}}(u)=
\begin{pmatrix}
1-u&&&\\
&1+u&&\\
&&1+u&\\
&&&1+u
\end{pmatrix}=\left(2Q_{\bar{0}}^{A_1^\dag}-1\right)u+1.\label{cs4}
\eea
It is not hard to verify that these are c-number solutions of (\ref{RE1})-(\ref{RE4}). Based on these solutions, the projection condition (\ref{projection}) for the regular part of the solution (\ref{singular}) has been checked in the Appendix~B.
We choose the scalar function $k(u)$ and $\bar{k}(u)$  in (\ref{kk1}, \ref{kk2}) to be
\bea
k(u)=\frac{1-u^2}{2u},\quad \bar{k}(u)=\frac{2-u}{2u}.
\eea
By plugging (\ref{Kminus1}), (\ref{Kminus2}), (\ref{Kplus1}) and (\ref{Kplus2}) obtained from the above solutions into (\ref{rbt}) and (\ref{lbt}) , we find
\bea\label{rrbt} \no
\mathcal{H}^{total}_r&=&\left(-2Q^{B_1}_{\overline{2L}}\left(\frac{1}{2}\mathbb{K}_{2L-1,\overline{2L}}-\mathbb{P}_{\overline{2L-2},\overline{2L}}+\frac{1}{2}\mathbb{P}_{\overline{2L-2},\overline{2L}}\mathbb{K}_{\overline{2L-2},2L-1}
+\frac{1}{2}\mathbb{K}_{\overline{2L-2},2L-1}\mathbb{P}_{\overline{2L-2},\overline{2L}}\right)Q^{B_1}_{\overline{2L}}\right.\\\no
&&\left.+2Q^{B^{\dg}_1}_{2L-1}Q^{B_1}_{\overline{2L}}+\frac{1}{2}\mathbb{K}_{\overline{2L-2},2L-1}Q^{B_1}_{\overline{2L}}\right)Q^{A_1}_1,
\eea
and
\bea\label{rlbt}
\mathcal{H}^{total}_l=\left(-2Q^{A_1}_1\left(\frac{1}{2}\mathbb{K}_{1\bar2}-\mathbb{P}_{13}+\frac{1}{2}\mathbb{P}_{13}\mathbb{K}_{1\bar2}+\frac{1}{2}\mathbb{K}_{1\bar2}\mathbb{P}_{13}\right)Q^{A_1}_1
+2Q^{A_1}_1Q^{A^{\dg}_1}_{\bar2}+\frac{3}{2}Q^{A_1}_1\right)Q^{B_1}_{\overline{2L}}.
\eea
Note that $\frac{1}2\mathbb{K}_{\overline{2L-2},2L-1}Q^{B_1}_{\overline{2L}}Q^{A_1}_1$ in $\mathcal{H}^{total}_r$ is actually a bulk term and will be cancelled by the related terms in
\bea
\mathcal{H}^{total}_{bulk}&=&\mathcal{H}_{bulk}+\bar{\mathcal{H}}_{bulk}\\\no
&=&-2 \sum_{l=1}^{L-2}\left(-\frac12\mathbb{I}-\mathbb{P}_{2l+1, 2l+3}+\frac12\mathbb{P}_{2l+1, 2l+3}\mathbb{K}_{2l+1, \overline{2l+2}}+\frac12\mathbb{P}_{2l+1, 2l+3}\mathbb{K}_{\overline{2l+2}, 2l+3}\right)
   Q^{A_1}_1Q^{B_1}_{\overline{2L}} \\ \no
   &-&2\sum_{l=1}^{L-2}\left(-\frac12\mathbb{I}-\mathbb{P}_{\overline{2l},\overline{2l+2}}+\frac{1}{2}\mathbb{P}_{\overline{2l},\overline{2l+2}}\mathbb{K}_{\overline{2l},2l+1}
   +\frac{1}{2}\mathbb{P}_{\overline{2l},\overline{2l+2}}\mathbb{K}_{2l+1,\overline{2l+2}}\right)Q^{A_1}_1Q^{B_1}_{\overline{2L}}\\\no
 &&+Q_1^{A_1}Q^{B_1}_{\overline{2L}}-\frac12\mathbb{K}_{2L-1, \overline{2L}}Q_1^{A_1}Q^{B_1}_{\overline{2L}}.\eea
Therefore, we see that, up to an overall factor ``$-\frac{\lambda^2}2$'' and some constant terms proportional to identity $\mathbb{I}$, we successfully reproduce the Hamiltonian (\ref{Hamiltonian}) from the integrable construction. Some calculation details are put in the Appendix~C.
\section{Bethe ansatz equations}\label{section4}
The commutativity property of $\tau(u)$ and $\bar\tau(u)$ implies the existence of $u$-independent eigenstates $\ket{\Lambda}$ of
both $\tau(u)$ and $\bar\tau(u)$,
\begin{equation}
\tau(u)\ket{\Lambda}=\Lambda(u)\ket{\Lambda},~~~\bar\tau(u)\ket{\Lambda}=\bar\Lambda(u)\ket{\Lambda}.
\end{equation}
We choose a reference state $\ket{(A_2B_2)^{L}}$ which is an eigenstate of both transfer matrices.
According to the direct calculations of $L=1,2,3$ cases, we conjecture that the eigenvalues of the reference state for a general $L$ are given by
\begin{equation}
\begin{split}\label{vac}
\Lambda_0(u)=\bar\Lambda_0(u)=&\frac{2}{u+1}\Big[-\frac{u^{2 L+2} (u+2)^{2 L}}{2 u+1}+\frac{(u+1)^{2 L+2} (u+2)^{2 L}}{2 u+1}\\&-\frac{(u+2)^{2 L+2} u^{2 L}}{2 u+3}+\frac{(u+1)^{2 L+2} u^{2 L}}{2 u+3}\Big]
\end{split}
\end{equation}
with appropriate rescaling of the transfer matrices by some functions of $u$ to cancel the denominators of $K^-_{2L-1, 2L}$ and $K^-_{2L-2, 2L}$. General eigenvalues should take a ``dressed'' form of (\ref{vac}).
Instead of the usual  methods based on algebraic Bethe ansatz or analytical Bethe ansatz, here we make a direct guess of the general eigenvalues.
The idea is to match the resulting Bethe ansatz equations to the coordinate Bethe ansatz given in \cite{Chen:2018sbp} for one-particle excitations.
We conjecture that general eigenvalues take the form
\begin{align}
 \Lambda(u|\{u_i\}) = & \frac{2(u+1)^{2L+1}(u+2)^{2L}}{2u+1}\frac{Q_1(u-\frac12)}{Q_1(u+\frac12)}
-\frac{2(u+2)^{2L}u^{2L+2}}{(2u+1)(u+1)}\frac{Q_1(u+\frac32)Q_2(u)}{Q_1(u+\frac12)Q_2(u+1)} \\
 & -\frac{2(u+2)^{2L+2}u^{2L}}{(2u+3)(u+1)}\frac{Q_2(u+2)Q_3(u+\frac12)}{Q_2(u+1)Q_3(u+\frac32)}
+\frac{2(u+1)^{2L+1}u^{2L}}{2u+3}\frac{Q_3(u+\frac52)}{Q_3(u+\frac32)},\\
 \bar\Lambda(u|\{u_i\})=  & \frac{2u^{2L}(u+1)^{2L+1}}{2u+3}\frac{Q_1(u+\frac52)}{Q_1(u+\frac32)}
 -\frac{2u^{2L}(u+2)^{2L+2}}{(2u+3)(u+1)}
\frac{Q_1(u+\frac12)Q_2(u+2)}{Q_1(u+\frac32)Q_2(u+1)}
\\
&-\frac{2u^{2L+2}(u+2)^{2L}}{(2u+1)(u+1)}\frac{Q_3(u+\frac32)Q_2(u)}{Q_3(u+\frac12)Q_2(u+1)}
+\frac{2(u+1)^{2L+1}(u+2)^{2L}}{2u+1}\frac{Q_3(u-\frac12)}{Q_3(u+\frac12)},
\end{align}

where
\be
Q_a(u)=\prod_{j=1}^{n_a}(u-iu_{a,j})(u+i u_{a,j}).
\ee
Notice that both $\Lambda(u)$ and $\bar\Lambda(u)$  are regular at $u=-1, -1/2, -3/2$. We also checked that their values at $u=0$ are consistent with (\ref{tau}) and (\ref{bartau}) taking into account the rescaling factors.
When taking $n_a=0, a=1, 2, 3$, they correctly reproduce the vacuum energy after carefully taking into account the rescaling factors and the constant term.

The analytic condition of $\Lambda(u)$  gives the Bethe ansatz equations
\begin{equation}
\begin{split}
e_1(u_{1,j})^{2L+2}=&\prod_{k=1}^{n_1}e_2(u_{1,j}-u_{1,k})e_2(u_{1,j}+u_{1,k})\prod_{k=1}^{n_2}e_{-1}(u_{1,j}-u_{2,k})e_{-1}(u_{1,j}+u_{2,k}),\\
-e_1(u_{2,j})e_{-2}^{2}(u_{2,j})=&\prod_{k=1}^{n_2}e_2(u_{2,j}-u_{2,k})e_2(u_{2,j}+u_{2,k})\prod_{k=1}^{n_1}e_{-1}(u_{2,j}-u_{1,k})e_{-1}(u_{2,j}+u_{1,k})
\times\\ &\times\prod_{k=1}^{n_3}e_{-1}(u_{2,j}-u_{3,k})e_{-1}(u_{2,j}+u_{3,k}),\\
e_1(u_{3,j})^{2L+2}=&\prod_{k=1}^{n_3}e_2(u_{3,j}-u_{3,k})e_2(u_{3,j}+u_{3,k})\prod_{k=1}^{n_2}e_{-1}(u_{3,j}-u_{2,k})e_{-1}(u_{3,j}+u_{2,k}),
\end{split}
\end{equation}
where
\begin{equation}
e_n(u)=\frac{u+\frac{in}{2}}{u-\frac{in}{2}}.
\end{equation}
As another consistency check, the analytic condition of $\bar\Lambda(u)$ gives the same equations.
Putting back the factor $-\frac{\lambda^2}2$ and subtracting a constant fixed by the fact that the vacuum has zero energy \cite{Chen:2018sbp}, the  eigenvalue of the Hamiltonian (\ref{Hamiltonian}) is given by
\be
E=\lambda^2\sum_{j=1}^{n_1}\frac{1}{u_{1,j}^2+\frac14}+\lambda^2\sum_{j=1}^{n_3}\frac{1}{u_{3,j}^2+\frac14}.
\ee
As further consistency checks, we compare our Bethe ansatz equations' solutions with eigenvalues by directly diagonalizing the transfer matrix $\tau(u)$ for some very simple cases. We define three Baxter polynomials for the corresponding single excitations
\be
Q_1(u)=(u-i u_1)(u+i u_1),Q_2(u)=(u-i u_2)(u+i u_2),Q_3(u)=(u-i u_3)(u+i u_3).
\ee
For $L=1$, the eigenvalues of $\tau(u)$ are proportional to\footnote{Notice that in \cite{Chen:2018sbp}, the finite size effects for operators with very small $L$ were not taken into account.}
\be
\{8,8+16u+8u^2,8-8u^2-4u^3,8+16u+16u^2+4u^3\}
\ee
where $8+16u+8u^2$ is triply degenerate and the others are doubly degenerate. All these eigenvalues can be reproduced by Bethe ansatz equations. For example, considering only $u_1$ excitation, the Bethe ansatz equation reads
\be
\left(\frac{u_1+\frac{i}{2}}{u_1-\frac{i}{2}}\right)^4=-\frac{2u_1+i}{2u_1-i}.
\ee
One solution is given by  $u_1=\frac{\sqrt{3}}{2}$,
\be
\Lambda(u|\{u_1=\frac{\sqrt{3}}{2}\})\big{|}_{L=1}=8-8u^2-4u^3,
\ee
while eigenvalue $8+16u+8u^2$ is reproduced by $\Lambda_0(u)|_{L=1}$. Other eigenvalues can be obtained by solving corresponding Bethe ansatz equations in a similar way.
Let's now study a slightly more complicated example. When $L=2$, we consider a single $A_1$ excitation
\be
\ket{\chi}=\ket{A_2B_2A_1B_2}.
\ee
The above state is an eigenstate of transfer matrix $\tau(u)$ with eigenvalue proportional to
\be
\chi=-4 (u+1)^2 \left(u^5+2 u^4-4 u^2-8 u-8\right)=\Lambda(u|\{u_1=\frac{\sqrt{3}}{2},u_2=-1\})\big{|}_{L=2}
\ee
where $u_1=\frac{\sqrt{3}}{2},u_2=-1$ can be obtained by solving the corresponding Bethe ansatz equations
\bea
\left(\frac{u_1+\frac{i}{2}}{u_1-\frac{i}{2}}\right)^6&=&-\frac{2u_1+i}{2u_1-i}\frac{u_1-u_2-\frac{i}{2}}{u_1-u_2+\frac{i}{2}}
\frac{u_1+u_2-\frac{i}{2}}{u_1+u_2+\frac{i}{2}},\\
-\frac{u_2+\frac{i}{2}}{u_2-\frac{i}{2}}\left(\frac{u_2-i}{u_2+i}\right)^2&=&-\frac{2u_2+i}{2u_2-i}\frac{u_2-u_1-\frac{i}{2}}{u_2-u_1+\frac{i}{2}}
\frac{u_2+u_1-\frac{i}{2}}{u_2+u_1+\frac{i}{2}}.
\eea

\section{Conclusion}\label{section5}
In this paper, we proved the integrability of the two-loop open spin chain Hamiltonian which stems from computing the anomalous dimension of determinant like operators in ABJM theory. The crucial part of the proof lies in the construction of projected operator-valued $K$-matrices. We also obtained the corresponding Bethe ansatz equations by guessing the eigenvalues of the double row transfer matrices directly avoiding  the complicated algebraic Bethe ansatz. Our ansatz of the eigenvalues of the transfer matrices passed through some consistency checks and should be compared with the all-loop asymptotic Bethe ansatz equations which need to be proposed. Such all-loop Bethe ansatz equations for  $\mathcal{N}=4$ SYM were obtained in \cite{Galleas:2009ye}.

We have shown in \cite{Chen:2018sbp} and this paper that determinant like operators provide integrable boundary conditions for the open spin chain in ABJM theory at least at planar two-loop level. Strong evidence for integrable open spin chain from flavored ABJM theory \cite{Hohenegger:2009as, Gaiotto:2009tk, Hikida:2009tp} was also found in \cite{Bai:2017jpe}. 
One of the further interesting  directions will be to investigate  whether some bosonic or fermionic BPS Wilson loops \cite{Drukker:2008zx}-\cite{Huang:2018mej} in ABJM theory could provide integrable boundary conditions as well.
The answer to this question in ${\mathcal N}=4$ SYM theory is positive \cite{Drukker:2006xg, Drukker:2012de, Correa:2012hh} and great success has been achieved.

\section*{Acknowledgments}
We would like to thank Nadav Drukker, Yunfeng Jiang and Gang Yang for very helpful  discussions.
The work of J.-B.~W.  was supported  by the National Natural Science Foundation of China, Grant No.\ 11575202.
The work of H.~O. was supported by the grant ``Exact Results in Gauge and String Theories'' from the Knut and Alice Wallenberg foundation.

\begin{appendix}\label{appendixa}
\section{Hamiltonian from transfer matrices}
Let us evaluate the Hamiltonian from $\frac{d}{du}\log \tau(u)$ and $\frac{d}{du}\log \bar{\tau}(u)$ at $u=0$ by starting with,
\bea\label{tau}
\tau(0)&=&\tr_0 K^{+}_{01}(0)[-2+\mathbb{K}_{0\bar2}]\mathbb{P}_{03}\cdots [-2+\mathbb{K}_{0, \overline{2L-4}}]\mathbb{P}_{0, 2L-3}\\\no
&&\times [-2+\mathbb{K}_{0, \overline{2L-2}}]{\mathbb{P}_{0, 2L-1}K^{-}_{0, \overline{2L}}(0)\mathbb{P}_{0, 2L-1}}[-2+\mathbb{K}_{0, \overline{2L-2}}]\\\no
&&\times \mathbb{P}_{0, 2L-3}[-2+\mathbb{K}_{0, \overline{2L-4}}]\cdots \mathbb{P}_{03}[-2+\mathbb{K}_{0\bar2}]\\\no
&=&2^{2(L-1)}[\tr_0 K^{+}_{01}(0)]K^{-}_{2L-1,\overline{2L}}(0),
\eea
\bea\label{bartau}
\bar{\tau}(0)&=&\tr_{\bar{0}}K^{+}_{\bar{0}1}(0)\mathbb{P}_{\bar{0}\bar2}[-2+\mathbb{K}_{\bar{0}3}]\cdots \mathbb{P}_{\bar{0}, \overline{2L-4}}[-2+\mathbb{K}_{\bar{0}, 2L-3}]\\\no
&&{\times \mathbb{P}_{\bar{0}, \overline{2L-2}}[-2+\mathbb{K}_{\bar{0}, 2L-1}]K^-_{\bar{0}, \overline{2L}}(0)[-2+\mathbb{K}_{\bar{0}, 2L-1}]\mathbb{P}_{\bar{0}, \overline{2L-2}}}\\\no
&&\times [-2+\mathbb{K}_{\bar{0}, 2L-3}]\mathbb{P}_{\bar{0}, \overline{2L-4}}\cdots [-2+\mathbb{K}_{\bar{0}3}]\mathbb{P}_{\bar{0}\bar2}\\\no
&=&2^{2(L-2)}[\tr_{\bar{0}}K^{+}_{\bar{0}1}(0)][-2+\mathbb{K}_{\overline{2L-2},2L-1}]K^-_{\overline{2L-2},\overline{2L}}(0)[-2+\mathbb{K}_{\overline{2L-2},2L-1}].
\eea

The derivative of $\tau(u)$ with respective to $\tau$ can be written as
\bea
\frac{d\tau(u)}{du}=\delta_l(u)+\sum_{i=1}^{L-1}\delta^i_1(u)+\sum_{i=1}^{L-1}\delta^i_2(u)+\delta_r(u)+\sum_{i=1}^{L-1}\delta^i_3(u)+\sum_{i=1}^{L-1}\delta^i_4(u), \eea
with
\bea \delta_l(u)\equiv \tr_0\left[\frac{d}{du}K^+_{01}(u)\right]R_{0\bar2}(u)\cdots R_{0, 2L-1}(u)K^{-}_{0, \overline{2L}}(u)R_{0,2L-1}(u)\cdots R_{0\bar2}(u), \eea

\bea
\delta^i_1(u)&\equiv&\tr_0K^{+}_{01}(u)R_{0\bar2}(u)R_{03}(u)\cdots \left[\frac{d}{du}R_{0, \overline{2i}}(u)\right]R_{0, 2i+1}(u)\cdots R_{0, \overline{2L-2}}(u)R_{0, 2L-1}(u)\\\no
&&\times K^{-}_{0,\overline{2L}}(u)R_{0,2L-1}(u)\cdots R_{0\bar2}(u),
\eea

\bea
\delta^i_2(u)&\equiv&\tr_0K^{+}_{01}(u)R_{0\bar2}(u)R_{03}(u)\cdots R_{0, \overline{2i}}(u)\left[\frac{d}{du}R_{0, 2i+1}(u)\right]\cdots R_{0, \overline{2L-2}}(u)R_{0, 2L-1}(u)\\\no
&&\times K^{-}_{0,\overline{2L}}(u)R_{0,2L-1}(u)\cdots R_{0\bar2}(u),
\eea

\bea
\delta_r(u) &\equiv& \tr_0K^{+}_{01}(u)R_{0\bar2}(u)\cdots R_{0, 2L-1}(u)\left[\frac{d}{du}K^{-}_{0, \overline{2L}}(u)\right]R_{0, 2L-1}(u)\cdots R_{0\bar2}(u),
\eea

\bea
\delta^i_3(u)&\equiv&\tr_0K^+_{01}(u)R_{0\bar2}(u)\cdots R_{0, 2L-1}(u)K^-_{0, \overline{2L}}(u)\\\no
&&\times R_{0, 2L-1}(u)\cdots \left[\frac{d}{du}R_{0, 2L-2i+1}(u)\right]R_{0, \overline{2L-2i}}(u)\cdots R_{0\bar2}(u),
\eea

\bea
\delta^i_4(u)&\equiv&\tr_0K^+_{01}(u)R_{0\bar2}(u)\cdots R_{0, 2L-1}(u)K^-_{0, \overline{2L}}(u)\\\no
&&\times R_{0,2L-1}(u)\cdots R_{0, 2L-2i+1}(u)\left[\frac{d}{du}R_{0, \overline{2L-2i}}(u)\right]\cdots R_{0\bar2}(u).
\eea

These terms can be simplified significantly at $u=0$,
\bea \delta_l(0)=2^{2(L-1)}[\tr_0 \frac{d}{du} K^{+}_{01}(u)\bigg|_{u=0}]K^{-}_{2L-1,\overline{2L}}(0),\eea

\bea
  \delta^1_{1}(0)=-2^{2(L-2)}\tr_0(K^{+}_{01}(0)(-2+\mathbb{K}_{0\bar2}))K^{-}_{2L-1,\overline{2L}}(0).
  \eea

When $1<i<L-1$, we have
\bea
  \delta^i_1(0)=-2^{2(L-2)}\tr_0K^{+}_{01}(0)K^{-}_{2L-1, \overline{2L}}(0)\left[-2+\mathbb{K}_{2i-1, \overline{2i}}\right],
  \eea

  \bea
  \delta^{L-1}_{1}(0)=-2^{2(L-2)}\tr_0K^{+}_{01}(0)K^{-}_{2L-1, \overline{2L}}(0)\left[-2+\mathbb{K}_{2L-3, \overline{2L-2}}\right],
  \eea

   \bea
   \delta^1_{2}(0)=2^{2(L-2)}\tr_0\left(K^+_{01}(0)\left[4\mathbb{P}_{03}-2\mathbb{P}_{03}\mathbb{K}_{0\bar2}-2\mathbb{K}_{0\bar2}\mathbb{P}_{03}+\mathbb{K}_{0\bar2}
   \right]\right)
   K^-_{2L-1,\overline{2L}}(0).
   \eea

   When $1<i<L-1$,
  \bea
  \delta^i_2(0)&=&2^{2(L-2)}\tr_0K^{+}_{01}(0)K^-_{2L-1,\overline{2L}}(0)\\\no
  &&\left[4\mathbb{P}_{2i-1, 2i+1}-2\mathbb{P}_{2i-1, 2i+1}\mathbb{K}_{2i-1,\overline{2i}}-2\mathbb{K}_{2i-1,\overline{2i}}\mathbb{P}_{2i-1, 2i+1}+\mathbb{K}_{2i-1, \overline{2i}}\right],
  \eea

  \bea
  \delta^{L-1}_{2}(0)&=&2^{2(L-2)}\tr_0K^{+}_{01}(0)K^-_{2L-1,\overline{2L}}(0)\\\no
  &&\left[4\mathbb{P}_{2L-3, 2L-1}-2\mathbb{P}_{2L-3,2L-1}\mathbb{K}_{2L-3, \overline{2L-2}}-2\mathbb{K}_{2L-3, \overline{2L-2}}\mathbb{P}_{2L-3, 2L-1}+\mathbb{K}_{2L-3, \overline{2L-2}}\right],
  \eea

\bea
\delta_r(0)=2^{2(L-2)}[\tr_0 K^{+}_{01}(0)]\left[\frac{d}{du}K^{-}_{2L-1, \overline{2L}}(u)\bigg|_{u=0}\right].
\eea

We also have\footnote{Here and in the following we have already used that $K^-_{2L-1, 2L}(0)=Q^{B_1}_{2L}$ obtained from (\ref{Kminus1}),
 and $\tr_0K^+_{01}(0)\propto Q_1^{A_1}$ from (\ref{Kplus1}). } \be \delta^i_3(0)=\delta^{L-i}_2(0),\, \delta^i_4(0)=\delta^{L-i}_1(0),\ee
for $1\le i \le L-1$.

So the contribution to the Hamiltonian from $\tau(u)$
\be\mathcal{H}=\tau(0)^{-1}\frac{d\tau(u)}{du}\bigg|_{u=0},\ee
can be decomposed as
\be\mathcal{H}=\mathcal{H}_l+\mathcal{H}_{bulk}+\mathcal{H}_r, \ee
with
\bea \mathcal{H}_l&=&\tau(0)^{-1}(\delta_l(0)+\delta^1_1(0)+\delta^1_2(0)+\delta^{L-1}_3(0)+\delta^{L-1}_4(0))\\ \no
&=&\left[\tr_0K^+_{01}(0)\right]^{-1}\left\{\tr_0\frac{dK^+_{01}(u)}{du}\bigg|_{u=0}+
\tr_0\left(K^+_{01}(0)\left(\mathbb{I}+2\mathbb{P}_{03}-\mathbb{P}_{03}\mathbb{K}_{0\bar2}-\mathbb{K}_{0\bar2}\mathbb{P}_{03}\right)\right)\right\}, \\
 \mathcal{H}_{bulk}&=&\tau(0)^{-1}\left(\sum_{i=2}^{L-1}\sum_{n=1}^2 \delta^i_a(0)+\sum_{i=1}^{L-2}\sum_{n=3}^4 \delta^i_a(0)\right) \\ \no
&=&\sum_{i=2}^{L-1}\left(2\mathbb{P}_{2i-1, 2i+1}-\mathbb{P}_{2i-1, 2i+1}\mathbb{K}_{2i-1, \overline{2i}}-\mathbb{K}_{2i-1, \overline{2i}}\mathbb{P}_{2i-1, 2i+1}+\mathbb{I}\right)Q_1^{A_1}Q^{B_1}_{\overline{2L}},\\
\mathcal{H}_r&=&\tau(0)^{-1}\delta_r(0)\\\no
&=&\left[K^{-}_{2L-1, \overline{2L}}(0)\right]^{-1}\left[\frac{d}{du}K^{-}_{2L-1, \overline{2L}}(u)\bigg|_{u=0}\right].
\eea

Similarly, we have \bea
\frac{d\bar{\tau}(u)}{du}=\bar{\delta}_l(u)+\sum_{i=1}^{L-1}\sum_{a=1}^4\bar{\delta}^i_a(u)+\bar{\delta}_r(u), \eea
with
\bea \bar{\delta}_l(u)\equiv \tr_{\bar{0}}\left[\frac{d}{du}K^+_{\bar{0}1}(u)\right]R_{\bar{0}\bar2}(u)\cdots R_{\bar{0},2L-1}(u)K^{-}_{\bar{0}, \overline{2L}}(u)R_{\bar{0}, 2L-1}(u)\cdots R_{\bar{0}\bar2}(u), \eea
\bea
\bar{\delta}^i_1(u)&\equiv&\tr_{\bar{0}}K^+_{\bar{0}1}(u)R_{\bar{0}\bar2}(u)\cdots\left[\frac{d}{du}R_{\bar{0}, \overline{2i}}(u)\right]R_{\bar{0}, 2i+1}(u)\cdots R_{\bar{0}, 2L-1}(u)\\\no
&&\times K^{-}_{\bar{0}, \overline{2L}}(u)R_{\bar{0}, 2L-1}(u)\cdots R_{\bar{0}\bar2}(u),
\eea

\bea
\bar{\delta}^i_2(u)&\equiv&\tr_{\bar{0}}K^+_{\bar{0}1}(u)R_{\bar{0}\bar2}(u)\cdots R_{\bar{0}, \overline{2i}}(u)\left[\frac{d}{du}R_{\bar{0}, 2i+1}(u)\right]\cdots R_{\bar{0}, 2L-1}(u)\\\no
&&\times K^{-}_{\bar{0}, \overline{2L}}(u)R_{\bar{0}, 2L-1}(u)\cdots R_{\bar{0}\bar2}(u),
\eea

\bea
\bar{\delta}_r(u)&\equiv& \tr_{\bar{0}}K^{+}_{\bar{0}1}(u)R_{\bar{0}\bar2}(u)\cdots R_{\bar{0}, 2L-1}\left[\frac{d}{du}{K^-_{\bar{0}, \overline{2L}}(u)}\right]R_{\bar{0}, 2L-1}(u)\cdots R_{\bar{0}\bar2}(u),\eea

\bea
\bar{\delta}^i_{3}(u)&\equiv&\tr_{\bar{0}}K^+_{\bar{0}1}(u)R_{\bar{0}\bar2}(u)\cdots R_{\bar{0}, 2L-1}(u)K^-_{\bar{0}, \overline{2L}}(u)\\\no
&&\times R_{\bar{0}, 2L-1}(u)\cdots \left[\frac{d}{du}R_{\bar{0}, 2L-2i+1}(u)\right]R_{\bar{0}, \overline{2L-2i}}(u)\cdots R_{\bar{0}\bar2}(u),
\eea

\bea
\bar{\delta}^i_{4}(u)&\equiv&\tr_{\bar{0}}K^+_{\bar{0}1}(u)R_{\bar{0}\bar2}(u)\cdots R_{\bar{0}, 2L-1}(u)K^-_{\bar{0}, \overline{2L}}(u)\\\no
&&\times R_{\bar{0}, 2L-1}(u)\cdots R_{\bar{0}, 2L-2i+1}(u)\left[\frac{d}{du}R_{\bar{0}, \overline{2L-2i}}(u)\right]\cdots R_{\bar{0}\bar2}(u).
\eea

We have that
\bea
\bar{\delta}_l(0)=2^{2(L-2)}[\tr_{\bar{0}}\frac{dK^{+}_{\bar{0}1}(u)}{du}\bigg|_{u=0}][-2+\mathbb{K}_{\overline{2L-2},2L-1}]K^-_{\overline{2L-2},\overline{2L}}(0)[-2+\mathbb{K}_{\overline{2L-2},2L-1}],
\eea

\bea
  \bar{\delta}^1_{1}(0)=2^{2(L-2)}\left[\tr_{\bar{0}}K^+_{\bar{0}1}(0)\mathbb{P}_{\bar{0}\bar2}\right]\left[-2+\mathbb{K}_{\overline{2L-2},2L-1}\right]K^-_{\overline{2L-2},\overline{2L}}(0)\left[-2+\mathbb{K}_{\overline{2L-2},2L-1}\right].
  \eea

  When  $1<i<L-1$,
  \bea
  \bar{\delta}^i_{1}(0)&=&2^{2(L-3)}\tr_{\bar{0}}K^+_{\bar{0}1}(0)\left[-2+\mathbb{K}_{\overline{2L-2},2L-1}\right]K^-_{\overline{2L-2},\overline{2L}}(0)\left[-2+\mathbb{K}_{\overline{2L-2},2L-1}\right]\\\no
  &&\times \left[4\mathbb{P}_{\overline{2i-2},\overline{2i}}-2\mathbb{P}_{\overline{2i-2},\overline{2i}}\mathbb{K}_{\overline{2i-2},2i-1}-2\mathbb{K}_{\overline{2i-2},2i-1}\mathbb{P}_{\overline{2i-2},\overline{2i}}+\mathbb{K}_{\overline{2i-2},2i-1}\right],
  \eea

   \bea
  \bar{\delta}^{L-1}_{1}(0)&=&2^{2(L-3)}\tr_{\bar{0}}K^+_{\bar{0}1}(0)\\\no
  &&\times \left[4\mathbb{P}_{\overline{2L-4},\overline{2L-2}}-2\mathbb{P}_{\overline{2L-4},\overline{2L-2}}\mathbb{K}_{\overline{2L-4},2L-3}-2\mathbb{K}_{\overline{2L-4},2L-3}\mathbb{P}_{\overline{2L-4},\overline{2L-2}}+\mathbb{K}_{\overline{2L-4},2L-3}\right]\\\no
  &&\times
  \left[-2+\mathbb{K}_{\overline{2L-2},2L-1}\right]K^-_{\overline{2L-2},\overline{2L}}(0)\left[-2+\mathbb{K}_{\overline{2L-2},2L-1}\right],
  \eea

  \bea
  \bar{\delta}^1_{2}(0)&=&-2^{2(L-3)}\tr_{\bar{0}}K^+_{\bar{0}1}(0)\left[-2+\mathbb{K}_{\overline{2L-2},2L-1}\right]K^-_{\overline{2L-2},\overline{2L}}(0)\left[-2+\mathbb{K}_{\overline{2L-2},2L-1}\right]\\\no
  &&\times \left[-2+\mathbb{K}_{\bar23}\right].
  \eea

 When $1<i<L-1$,
  \bea
  \bar{\delta}^i_2(0)&=&-2^{2(L-3)}\tr_{\bar{0}}K^+_{\bar{0}1}(0)\left[-2+\mathbb{K}_{\overline{2L-2},2L-1}\right]K^-_{\overline{2L-2},\overline{2L}}(0)\left[-2+\mathbb{K}_{\overline{2L-2},2L-1}\right]\\\no
  &&\times \left[-2+\mathbb{K}_{\overline{2i},2i+1}\right],
  \eea

  \bea
  \bar{\delta}^{L-1}_{2}(0)=-2^{2(L-2)}\tr_{\bar{0}}K^+_{\bar{0}1}(0)K^-_{\overline{2L-2},\overline{2L}}(0)\left[-2+\mathbb{K}_{\overline{2L-2},2L-1}\right],\label{cancel1}
  \eea

\bea\bar{\delta}_r(0)=2^{2(L-2)}[\tr_{\bar{0}}K^{+}_{\bar{0}1}(0)][-2+\mathbb{K}_{\overline{2L-2}, 2L-1}]\left[\frac{d}{du}K^-_{\overline{2L-2}, 2L-1}(u)\bigg|_{u=0}\right][-2+\mathbb{K}_{\overline{2L-2}, 2L-1}]. \eea
We also have\footnote{Here and in the following,  we have used  that $K^-_{\overline{2L-2}, \overline{2L}}(0)=Q^{B_1}_{\overline{2L}}$ obtained from  (\ref{Kminus2}),
 and $\tr_{\bar{0}}K^+_{\bar{0}1}(0)\propto Q_1^{A_1}$ from  (\ref{Kplus2}). }
\be \bar{\delta}^i_3(0)=\bar{\delta}^{L-i}_2(0),\, \bar{\delta}^i_4(0)=\bar{\delta}^{L-i}_1(0),\ee
for $1\le i \le L-1$.

The contribution to the Hamiltonian from $\bar{\tau}(u)$,
\be\bar{\mathcal{H}}=\bar{\tau}(0)^{-1}\frac{d\bar{\tau}(u)}{du}\bigg|_{u=0},\ee
can be decomposed as
\be\bar{\mathcal{H}}=\bar{\mathcal{H}}_l+\bar{\mathcal{H}}_{bulk}+\bar{\mathcal{H}}_r, \ee
with
\bea
\bar{\mathcal{H}}_{l}&=&\bar{\tau}(0)^{-1}\left[\bar{\delta}_l(0)+\bar{\delta}^1_{1}(0)+\bar{\delta}^{L-1}_{4}(0)\right]\\\no
&=&\left[\tr_{\bar{0}}K^+_{{\bar{0}}1}(0)\right]^{-1}\left(\tr_{\bar{0}}\frac{dK^+_{{\bar{0}}1}(u)}{du}\bigg|_{u=0}+
2\tr_{\bar{0}}\left(K^{+}_{\bar{0}1}(0)\mathbb{P}_{\bar{0}\bar2}\right)\right),\\
\bar{\mathcal{H}}_{bulk}&=&\bar{\tau}(0)^{-1}\left(\sum_{i=2}^{L-1}\delta^i_1(0)+\sum_{i=1}^{L-1}\delta^i_2(0)+\sum_{i=1}^{L-1}\delta^i_3(0)+\sum_{i=1}^{L-2}\delta^i_4(0)\right)\\ \no
 &=&\sum_{i=1}^{L-2}\left( 2\mathbb{P}_{\overline{2i}, \overline{2i+2}}-\mathbb{K}_{\overline{2i}, 2i+1}\mathbb{P}_{\overline{2i}, \overline{2i+2}}-\mathbb{P}_{\overline{2i}, \overline{2i+2}}\mathbb{K}_{\overline{2i}, 2i+1}+\mathbb{I}\right)Q_1^{A_1}Q^{B_1}_{\overline{2L}}\\ \no
 &&+Q_1^{A_1}Q^{B_1}_{\overline{2L}}-\frac12\mathbb{K}_{2L-1, \overline{2L}}Q_1^{A_1}Q^{B_1}_{\overline{2L}},\\
\bar{\mathcal{H}}_r&=&\bar{\tau}(0)^{-1}\bar{\delta}_r(0)\\\no
&=&[-2+\mathbb{K}_{\overline{2L-2},2L-1}]^{-1}\left[K^{-}_{\overline{2L-2},\overline{2L}}(0)\right]^{-1}\left[\frac{d}{du}K^{-}_{\overline{2L-2},\overline{2L}}(u)\bigg|_{u=0}\right][-2+\mathbb{K}_{\overline{2L-2},2L-1}].
\eea

Finally,
\bea \mathcal{H}^{total}&=&\mathcal{H}+\bar{\mathcal{H}}\\ \no
&=&\mathcal{H}^{total}_l+\mathcal{H}^{total}_{bulk}+\mathcal{H}^{total}_r, \eea
with
\bea
\mathcal{H}^{total}_l&=&\mathcal{H}_l+\bar{\mathcal{H}}_l\\\no
&=&\left[\tr_0K^+_{01}(0)\right]^{-1}\left\{\tr_0\frac{dK^+_{01}(u)}{du}\bigg|_{u=0}+
\tr_0\left(K^+_{01}(0)\left(\mathbb{I}+2\mathbb{P}_{03}-\mathbb{P}_{03}\mathbb{K}_{0\bar2}-\mathbb{K}_{0\bar2}\mathbb{P}_{03}\right)\right)\right\}\\\no
&+&\left[\tr_{\bar{0}}K^+_{{\bar{0}}1}(0)\right]^{-1}\left(\tr_{\bar{0}}\frac{dK^+_{{\bar{0}}1}(u)}{du}\bigg|_{u=0}+
2\tr_{\bar{0}}\left(K^{+}_{\bar{0}1}(0)\mathbb{P}_{\bar{0}\bar2}\right)\right),\\
\mathcal{H}^{total}_{bulk}&=&\mathcal{H}_{bulk}+\bar{\mathcal{H}}_{bulk}\\\no
&=& \sum_{l=1}^{L-2}\left(\mathbb{I}+2\mathbb{P}_{2l+1, 2l+3}-\mathbb{P}_{2l+1, 2l+3}\mathbb{K}_{2l+1, \overline{2l+2}}-\mathbb{P}_{2l+1, 2l+3}\mathbb{K}_{\overline{2l+2}, 2l+3}\right)
   Q^{A_1}_1Q^{B_1}_{\overline{2L}} \\ \no
   &+&\sum_{l=1}^{L-2}\left(\mathbb{I}+2\mathbb{P}_{\overline{2l},\overline{2l+2}}-\mathbb{P}_{\overline{2l},\overline{2l+2}}\mathbb{K}_{\overline{2l},2l+1}
   -\mathbb{P}_{\overline{2l},\overline{2l+2}}\mathbb{K}_{2l+1,\overline{2l+2}}\right)Q^{A_1}_1Q^{B_1}_{\overline{2L}}\\\no
 &&+Q_1^{A_1}Q^{B_1}_{\overline{2L}}-\frac12\mathbb{K}_{2L-1, \overline{2L}}Q_1^{A_1}Q^{B_1}_{\overline{2L}},\\
 \mathcal{H}^{total}_r&=&\mathcal{H}_r+\bar{\mathcal{H}}_r\\\no
&=&\left[K^{-}_{2L-1,\overline{2L}}(0)\right]^{-1}\left[\frac{d}{du}K^{-}_{2L-1,\overline{2L}}(u)\bigg|_{u=0}\right]\\\no
&+&[-2+\mathbb{K}_{\overline{2L-2},2L-1}]^{-1}\left[K^{-}_{\overline{2L-2},\overline{2L}}(0)\right]^{-1}\left[\frac{d}{du}K^{-}_{\overline{2L-2},\overline{2L}}(u)\bigg|_{u=0}\right][-2+\mathbb{K}_{\overline{2L-2},2L-1}].
\eea

\section{Projection condition for $K^-(u)$}\label{appendixb}
In this appendix, we will show that the following two regular solutions
\bea
&&\mathcal{K}^-_{1\bar2}(u)=R_{1\bar2}(u)K_1(u)R^{-1}_{1\bar2}(-u),\\\no
&&\mathcal{K}^-_{\bar1\bar2}(u)=R_{\bar1\bar2}(u)K_{\bar1}(u)R^{-1}_{\bar1\bar2}(-u),
\eea
satisfy the projection condition (\ref{projection}). By choosing the natural basis in the space $V_1$ ($V_{\bar{1}}$), we can express the $R$-matrices as an operator-valued matrix with  elements being the following
operators acting on the space $V_{\bar 2}$,
\bea
&&\left[R_{1\bar2}(u)\right]_{ij}=-(u+2)\delta_{ij}+|i\rangle\langle j|,\\\no
&&\left[R_{\bar1\bar2}(u)\right]_{ij}=u\delta_{ij}+|j\rangle\langle i|.
\eea
Recall the diagonal c-number solution $K_1(u)$ and $K_{\bar{1}}(u)$ given in (\ref{cs1}) and (\ref{cs2}),
\bea
&&\left[K_{1}(u)\right]_{ij}=\delta_{ij}h_i(u),\\\no
&&\left[K_{\bar{1}}(u)\right]_{ij}=\delta_{ij}\bar{h}_i(u),
\eea
with
\bea
h_i(u)=
\left\{
\begin{matrix}
1-u,\quad i=3\\1+u,\quad i\neq 3
\end{matrix}
\right.
;\qquad
\bar{h}_i(u)=
\left\{
\begin{matrix}
-1,\quad i=3\\1,\quad i\neq 3
\end{matrix}
\right..
\eea
Also note that the projectors at the right boundary are
\bea
Q^{B_1}_{\bar{2}}=I-|3\rangle\langle 3|,\quad \left(Q^{B_1}_{\bar{2}}\right)^{\perp}=|3\rangle\langle 3|.
\eea
Let us first focus on the regular solution $\mathcal{K}^-_{1\bar{2}}(u)$, the component can be easily found to be
\bea
\left[\mathcal{K}^-_{1\bar2}(u)\right]_{ij}=\frac{1}{4-u^2}\left[(u+2)^2\delta_{ij}h_i(u)+\left(\sum^4_{k=1}h_k(u)-(u+2)h_i(u)-(u+2)h_j(u)\right)|i\rangle\langle j|\right].
\eea
After the projection, we have
\bea
&&Q^{B_1}_{\bar{2}}\left[\mathcal{K}^-_{1\bar2}(u)\right]_{ij}\left(Q^{B_1}_{\bar{2}}\right)^{\perp}\\\no
&=&\frac1{4-u^2}\delta_{j3}\left(|i\rangle\langle 3|-\delta_{i3}|3\rangle\langle 3|\right)\left(\sum^4_{k=1}h_k(u)-(u+2)h_i(u)-(u+2)h_j(u)\right).
\eea
For $j\neq 3$ or $i=3$, the projection is automatically zero; otherwise for $j=3$ and $i\neq 3$, we have
\bea
\sum^4_{k=1}h_k(u)-(u+2)h_i(u)-(u+2)h_j(u)=(4+2u)-(u+2)(1+u)-(u+2)(1-u)=0.
\eea
So we see the projection condition is satisfied by the regular solution $\mathcal{K}^-_{1\bar{2}}(u)$. As for the second regular solution, we repeat the same analysis. The component of $\mathcal{K}^-_{\bar{1}\bar{2}}(u)$ is
\bea
\left[\mathcal{K}^-_{\bar{1}\bar{2}}(u)\right]_{ij}
=\frac{1}{1-u^2}\left[\left(u^2\bar{h}_i(u)+\sum_{k=1}^4\bar{h}_k(u)|k\rangle\langle k|\right)\delta_{ij}+u\left(\bar{h}_i(u)+\bar{h}_j(u)\right)|j\rangle\langle i|\right].
\eea
After the projection, we have
\bea
Q^{B_1}_{\bar{2}}\left[\mathcal{K}^-_{\bar{1}\bar{2}}(u)\right]_{ij}\left(Q^{B_1}_{\bar{2}}\right)^\perp
=\frac{u}{1-u^2}\delta_{i3}\left(|j\rangle\langle 3|-\delta_{j3}|3\rangle\langle 3|\right)\left(\bar{h}_i(u)+\bar{h}_j(u)\right).
\eea
For $i\neq 3$ or $j=3$, the projection is automatically zero; otherwise for $i=3$ and $j\neq 3$, we have
\bea
\bar{h}_i(u)+\bar{h}_j(u)=-1+1=0.
\eea
So the projection condition is satisfied as well. One may also check the projection condition for the regular solution contained in the projected $K^+$-matrix using totally the same procedure.

\section{Projected K-matrices from the given c-number solutions}\label{appendixc}
With the c-number solutions (\ref{cs1}) and (\ref{cs2}), we find the projected $K^-$-matrices are
\bea\label{Kminus1}
K^{-}_{2L-1, \overline{2L}}(u)&=&Q^{B_1}_{\overline{2L}}R_{2L-1, \overline{2L}}(u){K}_{2L-1}(u)R^{-1}_{2L-1, \overline{2L}}(-u)Q^{B_1}_{\overline{2L}}\\\no
&=&\frac{1}{2-u}\left[2u(u+2)Q^{B_1}_{\overline{2L}}Q^{B^{\dg}_1}_{2L-1}-2uQ^{B_1}_{\overline{2L}}\mathbb{K}_{2L-1,\overline{2L}}Q^{B_1}_{\overline{2L}}+(1-u)(u+2)Q^{B_1}_{\overline{2L}}\right]
\eea
and
\bea\label{Kminus2}
K^{-}_{\overline{2L-2}, \overline{2L}}(u)&=&Q^{B_1}_{\overline{2L}}R_{\overline{2L-2}, \overline{2L}}(u){K}_{\overline{2L-2}}(u)R^{-1}_{\overline{2L-2}, \overline{2L}}(-u)Q^{B_1}_{\overline{2L}}\\\no
&=&\frac{1}{1-u^2}\left[2u^2Q^{B_1}_{\overline{2L-2}}Q^{B_1}_{\overline{2L}}+2uQ^{B_1}_{\overline{2L}}\mathbb{P}_{\overline{2L-2},\overline{2L}}Q^{B_1}_{\overline{2L}}+(1-u^2)Q^{B_1}_{\overline{2L}}\right].
\eea
For the projected $K^+$-matrices, we fix the scalar function $k(u)$ and $\bar{k}(u)$  in (\ref{kk1}, \ref{kk2}) to be
\bea
k(u)=\frac{1-u^2}{2u},\quad \bar{k}(u)=\frac{2-u}{2u}.
\eea
Then, with the solutions ($\ref{cs3}$) and ($\ref{cs4}$), we find
\bea\label{Kplus1}
K^{+}_{01}(u)&=&k(u)\tr_a\mathbb{P}_{0a}R_{0a}(-2u-4)Q^{A_1}_1R_{1a}(u)\left(2Q^{A_1}_a-1\right)R^{-1}_{1a}(-u)Q^{A_1}_1\\\no
&=&\left(u^2+3u\right)Q^{A_1}_1-2u\left(u+2\right)Q^{A_1}_1Q^{A_1}_0-2\left(u+2\right)Q^{A_1}_1\mathbb{P}_{01}Q^{A_1}_1
\eea
and
\bea\label{Kplus2}
K^+_{\bar{0}1}(u)&=&\bar{k}(u)\tr_{\bar{a}}\mathbb{P}_{\bar{0}\bar{a}}R_{\bar{0}\bar{a}}(-2u-4)Q^{A_1}_1R_{1\bar{a}}(u)\left(2uQ^{A_1^{\dg}}_{\bar{a}}-u+1\right)R^{-1}_{1\bar{a}}(-u)Q^{A_1}_1\\\no
&=&\left(u^2+4u+3\right)Q^{A_1}_1-2\left(u+2\right)^2Q^{A^{\dg}_1}_{\bar{0}}Q^{A_1}_1+2\left(u+2\right)Q^{A_1}_1\mathbb{K}_{\bar{0}1}Q^{A_1}_1.
\eea
In deriving these quantities, we used the following several simple equalities
\bea
&&\tr_{\bar{2}} Q_{\bar{2}}^{A^\dg} \mathbb{K}_{1\bar{2}}=Q_1^A,\\
&&\mathbb{K}_{1\bar{2}}X_{\bar{2}}\mathbb{K}_{1\bar{2}}=\left(\tr X\right)\mathbb{K}_{1\bar{2}},\quad \forall X,\\
&&Q^{A}_1Q^{A^\dg}_{\bar{2}}\mathbb{K}_{1\bar{2}}Q^{A^\dg}_{\bar{2}}=Q^{A^\dg}_{\bar{2}}\mathbb{K}_{1\bar{2}}Q^{A^\dg}_{\bar{2}}Q^A_1=Q^{A^\dg}_{\bar{2}}\mathbb{K}_{1\bar{2}}Q^{A^\dg}_{\bar{2}},\\
&&Q^A_1Q^A_3\mathbb{P}_{13}Q^A_3=Q^A_3\mathbb{P}_{13}Q^A_1Q^A_3=Q^A_3\mathbb{P}_{12}Q^A_3,\\
&&Q^A_3\mathbb{P}_{13}Q^A_1\mathbb{P}_{13}Q^A_3=Q^A_3.
\eea
Then, substituting (\ref{Kminus1}), (\ref{Kminus2}), (\ref{Kplus1}) and (\ref{Kplus2}) into (\ref{rbt}) and (\ref{lbt}), we readily obtain the boundary Hamiltonian (\ref{rrbt}) and (\ref{rlbt}).

\end{appendix}

\end{document}